\documentclass[prb,eqsecnum,twocolumn,floatfix,groupedaddress]{revtex4}
\usepackage[dvips]{graphicx}
\usepackage{latexsym}
\usepackage[centertags]{amsmath}
\usepackage{amssymb}
\preprint{cond-mat/06mmxxx}

\newcommand{\ej}[1]{\eta_{\mu,{#1}}}
\newcommand{\tj}[1]{\Theta_{\mu,{#1}}}
\newcommand{\pj}[1]{|\Phi_{\mu,{#1}}|}
\newcommand{\dr}[1]{\delta\rho_{#1}}
\renewcommand{\vec}[1]{\boldsymbol{#1}}

\begin{document}
\title{From stripe to checkerboard order on the square lattice\\ in the presence of
quenched disorder}

\author{Adrian Del Maestro}
\affiliation{Department of Physics, Harvard University, Cambridge,
MA 02138}
\author{Bernd Rosenow}
\thanks{On leave from the Institut f\"ur Theoretische Physik, Universit\"at zu
K\"oln, D-50923, Germany} \affiliation{Department of Physics,
Harvard University, Cambridge, MA 02138}
\author{Subir Sachdev}
\affiliation{Department of Physics, Harvard University, Cambridge,
MA 02138}

\begin{abstract}
We discuss the effects of quenched disorder on a model of charge
density wave (CDW) ordering on the square lattice.  Our model may be
applicable to the cuprate superconductors, where a random
electrostatic potential exists in the CuO$_2$ planes as a result of
the presence of charged dopants. We argue that the presence of a
random potential can affect the unidirectionality of the CDW order,
characterized by an Ising order parameter. Coupling to a
unidirectional CDW, the random potential can lead to the formation
of domains with 90 degree relative orientation, thus tending to
restore the rotational symmetry of the underlying lattice. We find
that the correlation length of the Ising order can be significantly
larger than the CDW correlation length. For a checkerboard CDW on
the other hand, disorder generates spatial anisotropies on short
length scales and thus some degree of unidirectionality. We quantify 
these disorder effects and suggest new techniques for analyzing the 
local density of states (LDOS) data measured in scanning tunneling 
microscopy experiments.
\end{abstract}
\pacs{74.20.De, 74.62.Dh, 74.72.-h}
\maketitle

\section{Introduction}
\label{sec:intro}

One of the major stumbling blocks preventing a quantitative
confrontation between theory and experiment in the cuprate
superconductors is the influence of quenched disorder on the
experimental observations. The dopant ions exert a significant
electrostatic potential on the CuO$_2$ plane, and so unless the ions
can be carefully arranged in a regular pattern, the mobile charge
carriers experience a random potential. Recent STM observation
\cite{hanaguri,fang,ali,mcelroy,hashimoto} clearly display that
quenched randomness is crucial in determining the spatial
modulations of the local density of states.

There has been much recent interest in determining the nature of the
spin and charge density wave order (CDW) observed in STM, neutron,
and X-ray scattering in a variety of cuprate compounds at low
temperatures
\cite{hanaguri,fang,ali,mcelroy,hashimoto,Tranquada+97,Lee+99,tranquada+04,abbamonte}.
The quenched disorder acts on the CDW order as a ``random field'',
which is always a relevant perturbation at low temperatures: true
long-range order is disrupted at any finite random field strength
\cite{apy}. Nevertheless, one might hope that an analytic treatment
may be possible in the limit of weak random fields. Many such
analyses\cite{apy,natter90,giam,dsf,nattermann} have been carried out in the
literature, describing states with power-law correlations and
suppressed dislocations (or related topological defects) at
intermediate length scales. At the longest scales, dislocations
always proliferate and all correlations are expected to decay
exponentially; no analytic treatment is possible in this strong
coupling regime. As we will discuss below, current experiments on
the cuprates are in a regime dominated by dislocations, and there
does not appear to be any significant regime of applicability of the
defect-free theory. Consequently, we are forced to rely on numerical
simulations for an understanding of experiments. We will present
numerical results over a representative range of parameters. Our aim
is to allow insights into the underlying theory by a comparison of
experimental and numerical results.

\section{Model}
\label{sec:model}

A previous work by two of the authors \cite{delmaestro} studied the
influence of thermal fluctuations on density wave order on the
square lattice. Here, we will study the influence of quenched
randomness on the same underlying theory. A generic density was
defined which could be any observable invariant under spin rotations
and time reversal
\begin{equation}
\delta \rho (\vec{r}) = \mbox{Re} \left[ \Phi_x \mathrm{e}^{i
\vec{K}_x \cdot \vec{r}}\right] + \mbox{Re} \left[ \Phi_y
\mathrm{e}^{i \vec{K}_y \cdot \vec{r}} \right], \label{eq:rho}
\end{equation}
where ${\bf K}_x = (2 \pi /a) (1/p, 0)$, ${\bf K}_y = (2 \pi /a)
(0,1/p)$, and $\Phi_{x,y}$ are complex order parameters which were
assumed to vary slowly on the scale of a lattice spacing.

If both amplitudes $|\Phi_{x,y}|$ have nonzero expectation values,
the charge density is modulated in both $x$- and $y$-directions and
describes a solid on the square lattice. In addition, if the wave
length of the charge ordering is commensurate with the underlying
crystal, i.e.~for integer $p$, the density displays true long range
order. For incommensurate charge order, fluctuations due to finite
temperature will cause quasilong--range order with a power law decay
of correlation functions. If only one of the two amplitudes
$|\Phi_{x,y}|$ has nonzero expectation value, the density
Eq.~(\ref{eq:rho}) describes unidirectional (striped) CDW order.
Again, the presence of a commensurate lattice potential makes the
order long ranged at finite temperatures, whereas in the
incommensurate case it is quasilong--ranged.

\begin{figure}[t]
\centering
\includegraphics*[width=3.2in]{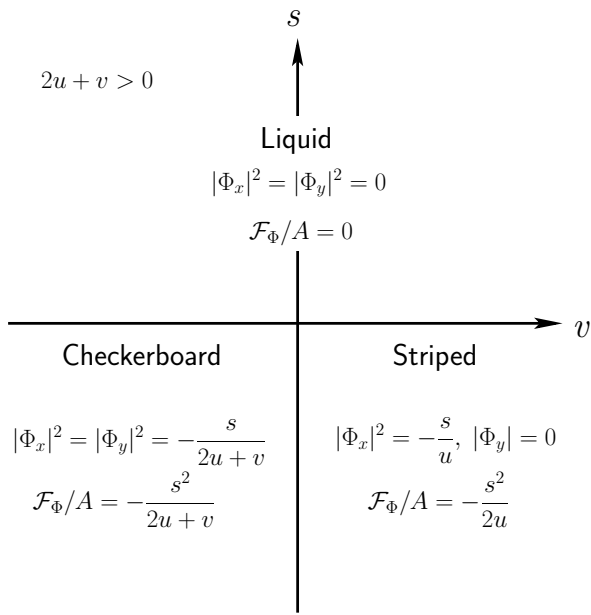}
\caption{The homogeneous mean field solutions of Eq.~(\ref{eq:FPhi})
for $\Phi_{x,y}$.} \label{fig:dfmft}
\end{figure}

In the incommensurate phase, in the absence of disorder, the most
general free energy density expanded in powers of $\Phi_{x,y}$ and
its gradients consistent with the symmetries of the square lattice
is given by
\begin{eqnarray}
\mathcal{F}_{\Phi} &=& \int d^2 r \Bigl[ C_1 \left( \left|
                       \partial_x \Phi_x \right|^2 + \left| \partial_y \Phi_y \right|^2
                       \right) \nonumber \\
                   &~& +\ C_2 \left( \left| \partial_y \Phi_x \right|^2 + \left|
               \partial_x \Phi_y \right|^2 \right)
                   +\ C_3  \left| \partial_x \Phi_x \right| \left|
               \partial_y \Phi_y \right|  \nonumber \\
           &~&~~~~~~+\ s \left( |\Phi_x |^2 + |\Phi_y |^2 \right) +
                    \frac{u}{2} \left(|\Phi_x |^2 + |\Phi_y |^2 \right)^2 \nonumber \\
           &~&~~~~~~+\ v |\Phi_x |^2 |\Phi_y |^2 \Bigr].
\label{eq:FPhi}
\end{eqnarray}
The homogeneous mean field solution of this model is summarized in
Fig.~\ref{fig:dfmft} where the checkerboard, stripe and liquid phase
values of $\Phi_{x,y}$ are shown along with the accompanying free
energy densities.

In this work we focus on the influence of quenched disorder on CDW
order and thus consider adding a term to the free energy consisting
of two complex random fields, coupling directly to $\Phi_{x,y}$
\begin{equation}
\mathcal{F}_{\mathrm{H}} = - \int d^2 r \left(\mathrm{H}_x^*\Phi_x +
\mathrm{H}_y^*\Phi_y + c.c. \right)
\label{eq:FH}
\end{equation}
resulting in the total action
\begin{equation}
\mathcal{F} = \mathcal{F}_\Phi + \mathcal{F}_\mathrm{H} \label{eq:F}
\end{equation}
where the complex random fields $\mathrm{H}_\mu$ ($\mu=x,y$) are
parameterized as $\mathrm{H}_\mu (\vec{r}) = h_\mu(\vec{r})
\mathrm{e}^{i\eta_\mu(\vec{r})}$, $h_\mu$ are Gaussian distributed
random variables with mean $0$, and standard deviation $h_0$ and
$\eta_\mu$ are uniformly distributed random phases on $[0,2\pi)$.

Let us confine ourselves to the condensed phase where $s < 0$, and
$u > 0$ for stability. The interesting physics are encapsulated by the
effects of altering the coupling constant $v$ and the variance of
the random field $h_0$. While $v$ changes the low energy ground
states from checkerboard like configurations for $v < 0$ ($|\Phi_x|
= |\Phi_y|$) to stripe like patterns ($|\Phi_\mu|\ne|\Phi_\nu| = 0$)
for $ v > 0$, the variance $h_0$ should destabilize both types of 
states.

A careful treatment of the coupling between CDW order and a random
electrostatic potential yields random compression terms of the form
\cite{nattermann} $\mathrm{h}_\mu \partial \Theta_\mu$ ($\Theta_\mu
= \mathrm{arg}[\Phi_\mu]$) omitted in the free energy
Eq.~(\ref{eq:FH}). In addition, an RG analysis of the full action
$\mathcal{F}_\Phi + \mathcal{F}_\mathrm{H}$ generates random shear
terms. Random compression and shear terms are responsible for the
power law decay of correlation functions on intermediate length
scales, on which the influence of topological excitations
(dislocations) in the phase fields $\Theta_\mu$ can be
neglected\cite{nattermann}.

In STM experiments, the correlation length of charge order is found
to have values ranging from\cite{hanaguri} 2.5, to roughly\cite{ali}
5 CDW periods.  In neutron scattering experiments, peak widths
corresponding to correlation lengths larger then ten CDW periods
were observed \cite{Tranquada+97,Lee+99}.  The correlation length
describes the scale on which dislocations proliferate, and the
presence of a relatively short correlation length indicates that
there is no intermediate length scale on which compression and shear
terms are important. For this reason, the omission of these terms
from the elastic energy should be justified.


\section{Numerical Minimization}
\label{sec:nummin}

Due to the presence of the two complex random fields
$\mathrm{H}_\mu(\vec{r})$, we have elected to minimize
Eq.~(\ref{eq:F}) numerically.  The interplay of elastic and disorder
energy causes frustration and gives rise to an exponentially large
number of low lying states with similar energies but very different
configurations. As these states are separated by large energy
barriers, relaxation after an external perturbation is very slow and
glassy dynamics can be observed. For these reasons, numerically
finding the ground state of such a system is a hard problem and as
novel algorithms are developed and employed, even to relatively
simple models, new states with lower energies are inevitably found
\cite{weigel}.

Gradient methods which move strictly downhill in the energy
landscape are fast, but prone to becoming stuck in local minima and
are not always able to reproduce the results of slower ergodic
methods.  Simulated annealing algorithms \cite{KiGeWe83} have been
the most successful at thoroughly sampling the possible
configuration space by using a fictitious temperature.  By
successively lowering this temperature, the resolution of finer and
finer energy scales becomes possible while avoiding the danger of
being stuck in a metastable excited state.

As a compromise we have elected to employ a combination of both
greedy conjugate gradient \cite{shewchuk} and ergodic simulated
annealing \cite{siarry} methods.  We allow for the possibility of
local uphill moves where the configuration update involves making a
downhill step in a random area of the sample. The size of the
randomly chosen region is annealed by tracking metropolis acceptance
rate. This method faithfully reproduces quite quickly, the results
of early Monte Carlo work on the random field XY model
\cite{gingras}.

We have performed simulations for a number of lattice sizes $L =
\{20,32,48,64,100\}$, commensurate with the experimentally observed
\cite{hanaguri,ali,fang,mcelroy} period of modulations in the local
density of states of four lattice spacings ($p = 4$) and multiple
realizations of disorder $N_{rd}(L) = \{200,200,150,150,100\}$.

Let us consider a $L \times L$ square lattice of $N$ sites labelled
by $i$, then after rescaling to give dimensionless coupling
parameters the continuum action Eq.~(\ref{eq:F}) in units where the
lattice constant $a_0$ is set to unity and $C_3 = 0$ takes the form
(with $\tj{i} = \mathrm{arg}[\Phi_{\mu,i}]$)
\begin{eqnarray}
\mathcal{F}_{L} &=& -\frac{1}{2} \sum_{\mu=x,y}\sum_{\langle i,j
\rangle} J_{\mu,j} \pj{i}\pj{j}
    \cos(\tj{i}-\tj{j}) \nonumber \\
    && +\ \sum_{\mu=x,y}\sum_{i}\Bigl[ (C_1 + C_2 + s)\pj{i}^2 + \frac{u}{2}\pj{i}^4
    \nonumber \\
    && -\ 2h_{\mu,i}\pj{i}\cos(\tj{i}-\ej{i}) \Bigr] \nonumber \\
    && +\ (u+v)\sum_i |\Phi_{x,i}|^2|\Phi_{y,i}|^2
\label{eq:FL}
\end{eqnarray}
with $\langle i,j \rangle $ indicating the usual sum over nearest
neighbors and the factor of $1/2$ is inserted to avoid double
counting. The coupling matrix $J_{\mu,j}$ has diagonal elements
$J_{x,i\pm x} = J_{y,i\pm y} = C_1$ and off diagonal couplings
$J_{x,i\pm y} = J_{y,i\pm x} = C_2$.  We have chosen to set $C_1 =
C_2 = 1$ and $s = -0.1$ thus restoring full rotational symmetry of
the elastic energy on scales much larger than the lattice spacing
and confining our analysis to the condensed phase.
We have also elected to pick the value of the quartic coupling $u$
to ensure the condensation energy remains constant across the
critical line $v = 0$ by setting $u(v\ge0) = -s$ and $u(v<0) = -(s + v/2)$.


\section{Results}
\label{sec:results}

\begin{figure}[t]
\centering
\includegraphics*[width=3.2in]{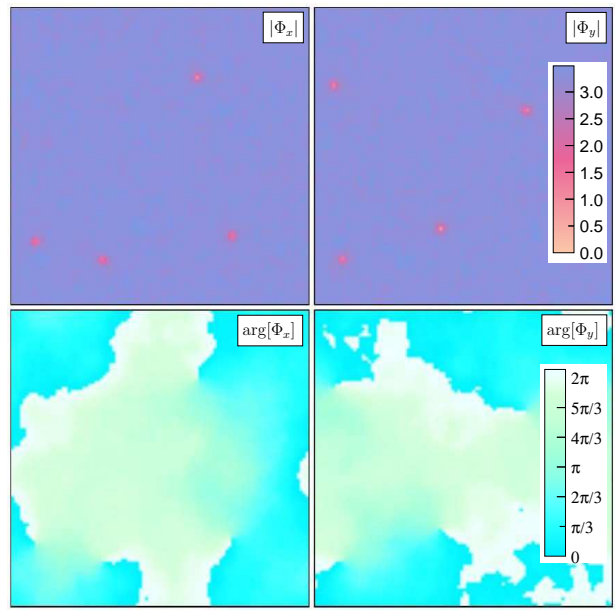}
\caption{The ground state field configuration on a $100\times 100$ lattice 
         for a random field standard deviation of $h_0 = 0.6$ and
	 $xy$-interaction $v = -0.1$ for a particular realization of disorder.}
\label{fig:OP1}
\end{figure}

\begin{figure}[t]
\centering
\includegraphics*[width=3.2in]{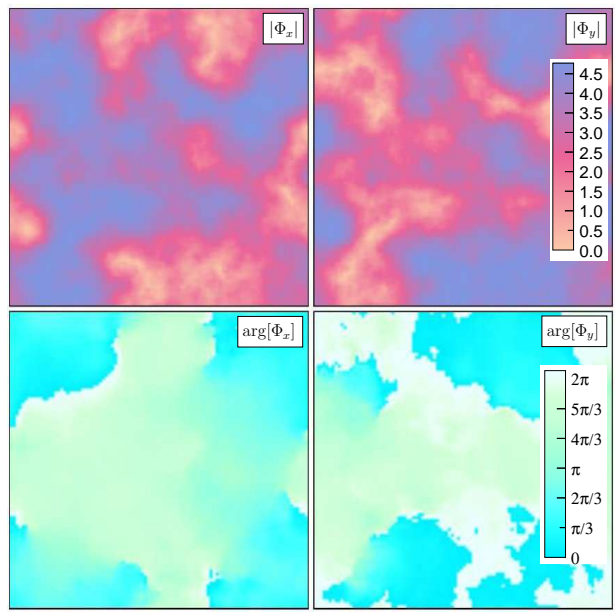}
\caption{The ground state field configuration on a $100\times 100$ lattice 
         for a random field standard deviation of $h_0 = 0.6$ and
	 $xy$-interaction $v = 0.0$ for a particular realization of disorder.}
\label{fig:OP2}
\end{figure}

\begin{figure}[t]
\centering
\includegraphics*[width=3.2in]{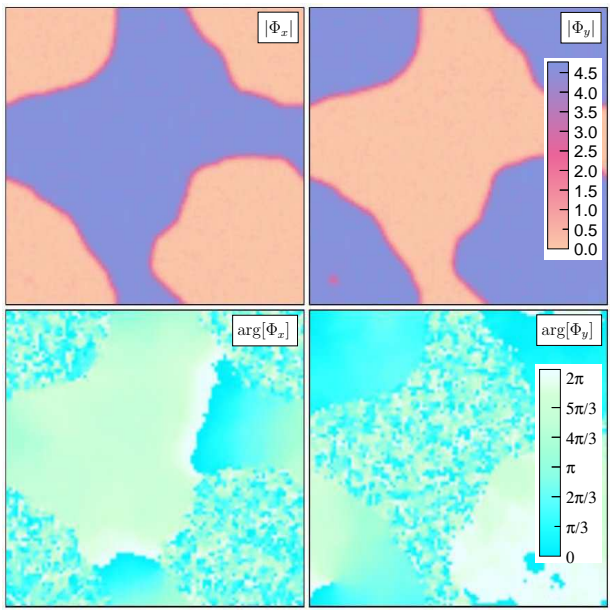}
\caption{The ground state field configuration on a $100\times 100$ lattice 
         for a random field standard deviation of $h_0 = 0.6$ and
	 $xy$-interaction $v = 0.1$ for a particular realization of disorder.
	 The phase of $\Phi_{x,y}$ fluctuates strongly in regions where its
	 amplitude is suppressed.}
\label{fig:OP3}
\end{figure}

Employing this minimization procedure we obtain stable low energy
field configurations like the ones shown in Figs.~\ref{fig:OP1} to 
\ref{fig:OP3}.
For $v = -0.1$ and $h_0 = 0.6$ (Fig.~\ref{fig:OP1}) we observe small circular 
regions less then 5 lattice spacings across,  where the local random field 
configuration has taken a value that makes it energetically favorable to 
suppress the magnitude of either $\Phi_x$ or $\Phi_y$ in its vicinity. 
For the uncoupled theory (Fig.~\ref{fig:OP2}) $\Phi_{x,y}$ appear much smoother
and although there still exists separate regions of stripe and checkerboard
order, the interfaces between these regions are poorly defined.
We may also observe the effects of positive $v = 0.1$ at the same disorder
strength as seen in Fig.~\ref{fig:OP3}.  In this case, the magnitude of 
either $\Phi_x$ or $\Phi_y$ is suppressed to zero over large regions of the 
sample leading to a ground state field configuration with well defined
domains having unidirectional order in either the $x$- or $y$-directions.

\begin{figure*}[t]
\centering \includegraphics*[width=7.0in]{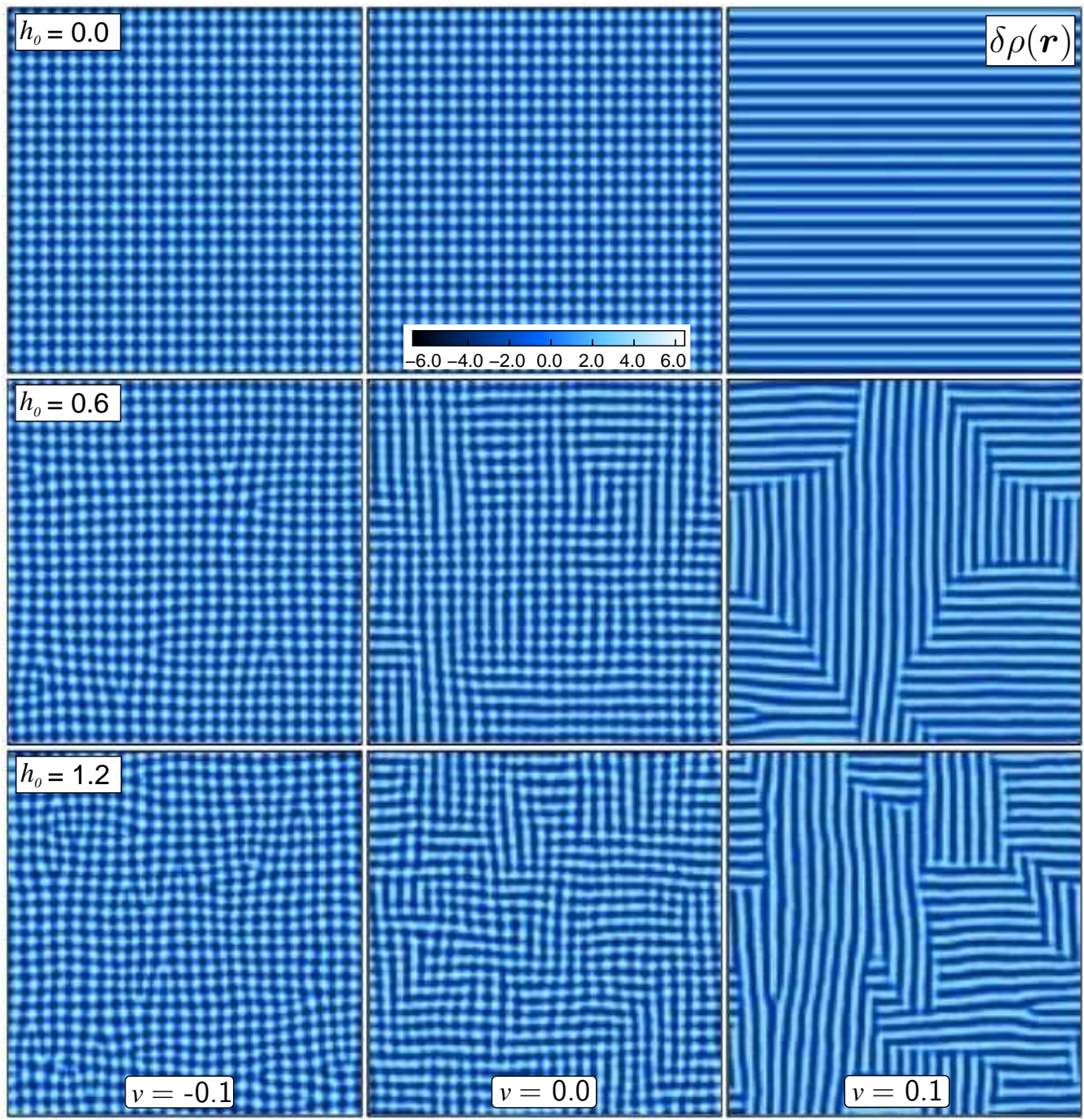}
\caption{The density fluctuations $\delta \rho(\vec{r})$ on a
  $100\times100$ lattice for $v = -0.1$ (left column), $v = 0.0$
  (center column) and $v = 0.1$ (right column) with 
  $h_0 = 0.0, 0.6$ and $1.2$ from top to bottom for one random field 
  configuration. The central row was constructed using the values
  of $\Phi_{x,y}$ from Figs.~\ref{fig:OP1} through \ref{fig:OP3}.}
\label{fig:drhovh}
\end{figure*}

We can construct the form of the density fluctuations $\delta
\rho(\vec{r})$ (Eq.~(\ref{eq:rho})) from the minimized value of the
spatially dependent order parameters to determine the effects of
altering $v$ and $h_0$ as seen in Fig.~\ref{fig:drhovh}.
For $v < 0$ and $h_0 = 0.0$ we observe robust checkerboard ordering
which is coherent over the entire lattice.  As the variance of the
random field is increased, dislocations in the phase of the
$\Phi_\mu$ fields gradually destroy the local ordering, and the
correlation length is reduced to less than three CDW periods for
$h_0 = 1.2$. For $v = 0.0$ and $h_0 = 0.0$, the mean field critical
value in the disorder free theory, density fluctuations in the $x$-
and $y$-direction have a period of four lattice spacings and
identical amplitudes.  However, the response of the system to
increasing $h_0$ is very different than in the $v=-0.1$ case.  For
any finite amount of disorder, there is a significant enhancement in
the size of unidirectional correlations as the sample breaks up into
regions with the magnitude of \emph{either} $\Phi_x$ or $\Phi_y$
greatly reduced. In the clean limit with $v > 0$ we obtain purely 
unidirectional density fluctuations.  As the strength of disorder is
increased, large domains of $\pi/2$ relatively orientated appear and finally
for large values of $h_0$ the system breaks up into many such regions of
varying sizes.

The presence of such strong stripeness leads to the natural
definition of a local Ising-like order parameter \cite{krmp}
\begin{equation}
\Sigma (\vec{r}) =  \frac{|\Phi_x(\vec{r})|^2 - |\Phi_y(\vec{r})|^2}
                         {|\Phi_x(\vec{r})|^2 + |\Phi_y(\vec{r})|^2}
\label{eq:sigma}
\end{equation}
which measures the tendency of the system to have only \emph{one} of
either $\Phi_x$ or $\Phi_y$ nonzero over some finite area, with a
value between -1 and 1.

As a result of our direct minimization procedure, we have calculated
a full set of low energy field configurations for multiple system
sizes and realizations of disorder with $xy$-couplings $v =
\{-0.1,0.0,0.1,0.2\}$ and random field variances $h_0$ between $0.0$
and $2.0$.  Using this information we can construct the disorder
averaged correlation functions for each system size throughout the
relevant parameter space.  Two distinct types of correlations are of
interest.  The first are simple CDW correlation functions between
the $\Phi_\mu$ fields given by
\begin{equation}
G_\mu (r) = \overline{\left \langle \Phi_\mu(r)
\Phi^{*}_\mu(0)\right\rangle} \label{eq:Gmu}
\end{equation}
where $\mu \in \{x,y\}$ while the second type measures fluctuations
of the Ising-like order parameter Eq.~(\ref{eq:sigma})
\begin{equation}
G_\Sigma (r) = \overline{\left \langle \Sigma(r) \Sigma(0) \right
\rangle} \label{eq:Gsigma}
\end{equation}
with $\langle \cdots \rangle$ indicating a spatial average in the
ground state and the over-line denotes an average over multiple
realizations of disorder $N_{rd}(L)$. As $L$ becomes large both
$\langle\Phi_{\mu}(r)\rangle$ and $\langle\Sigma(r)\rangle$ tend to zero,
and the connected and disconnected correlation functions are
equivalent. Note that the definition of $G_\Sigma (r)$ distinguishes
between regions with strong unidirectional order with $\pi/2$
relative orientation.

\begin{figure*}[t]
\centering
\includegraphics*[width=7.0in]{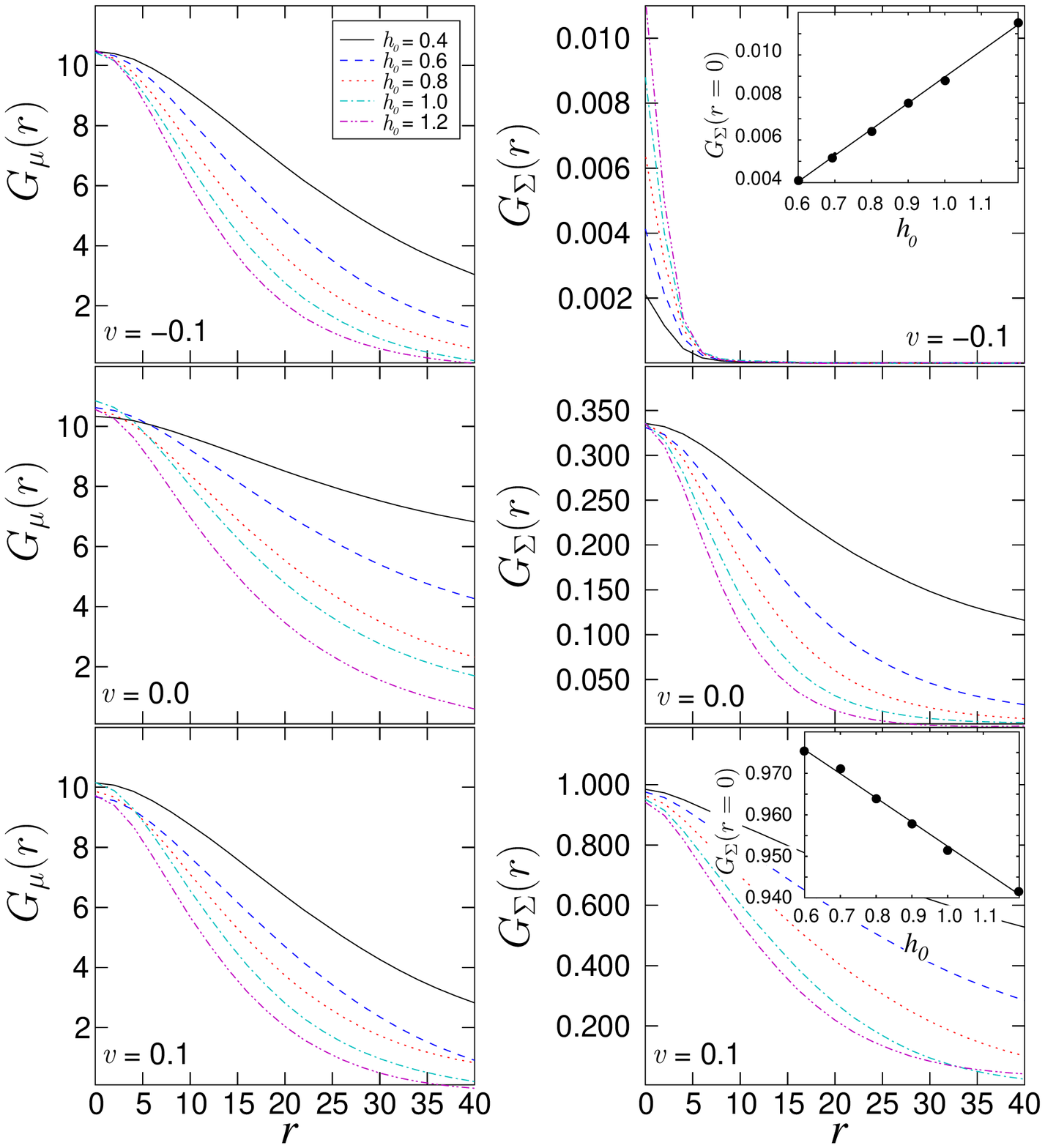}
\caption{Selected $G_\mu$ and $G_\Sigma$ correlation functions for
     $v = -0.1,0.0,0.1$ and $h_0 = 0.4,0.6,0.8,1.0,1.2$ (legend applies to all
     panels) which were averaged over 100 realizations of disorder in a
     $100\times100$ sample. The insets show
     $\overline{\left\langle\Sigma^2(0)\right\rangle}$ vs. $h_0$ for $v = -0.1$
         (upper right) and $v = 0.1$ (lower right).}
\label{fig:G}
\end{figure*}

For sufficiently large variances in the magnitude of the random
field, we expect that both Eqs.~(\ref{eq:Gmu}) and (\ref{eq:Gsigma})
will be characterized by exponential decays of the form
\begin{eqnarray}
G_\mu(r) &\sim& \mathrm{e}^{-r/\xi_\mu} \nonumber \\
G_\Sigma(r) &\sim& \mathrm{e}^{-r/\xi_\Sigma}
\end{eqnarray}
where $\xi_\mu$ and $\xi_\Sigma$ are their respective correlation
lengths.  $G_\mu(r)$ and $G_\Sigma(r)$ are shown for distances up to
40 lattice spacings for $v = -0.1,0.0$ and $0.1$ in
Fig.~\ref{fig:G}.

All correlations appear to decay exponentially for $h_0 \ge 0.4$,
and the most striking differences between $G_\mu(r)$ and
$G_\Sigma(r)$ can be seen at $r = 0$ by comparing
$\overline{\left\langle\Phi_\mu^2(0)\right\rangle}$ and
$\overline{\left\langle\Sigma^2(0)\right\rangle}$. The
$xy$-interaction parameter $v$ has little effect on the scale of the
background CDW order, while the background Ising-like order,
measured by $G_\Sigma(0)$, increases by three orders of magnitude as
$v$ changes from $-0.1$ to $0.1$.  In addition it appears that
(after proper finite size scaling) $G_\mu(0)$ is essentially a
monotonically decreasing function of $h_0$ from the checkerboard to
stripe parameter regime. The two insets in Fig.~\ref{fig:G} clearly
show however, that the slope of $G_\Sigma(0)$ vs. $h_0$ changes from
positive to negative near $v = 0.0$.

\begin{figure}[t]
\centering \includegraphics*[width=3.2in]{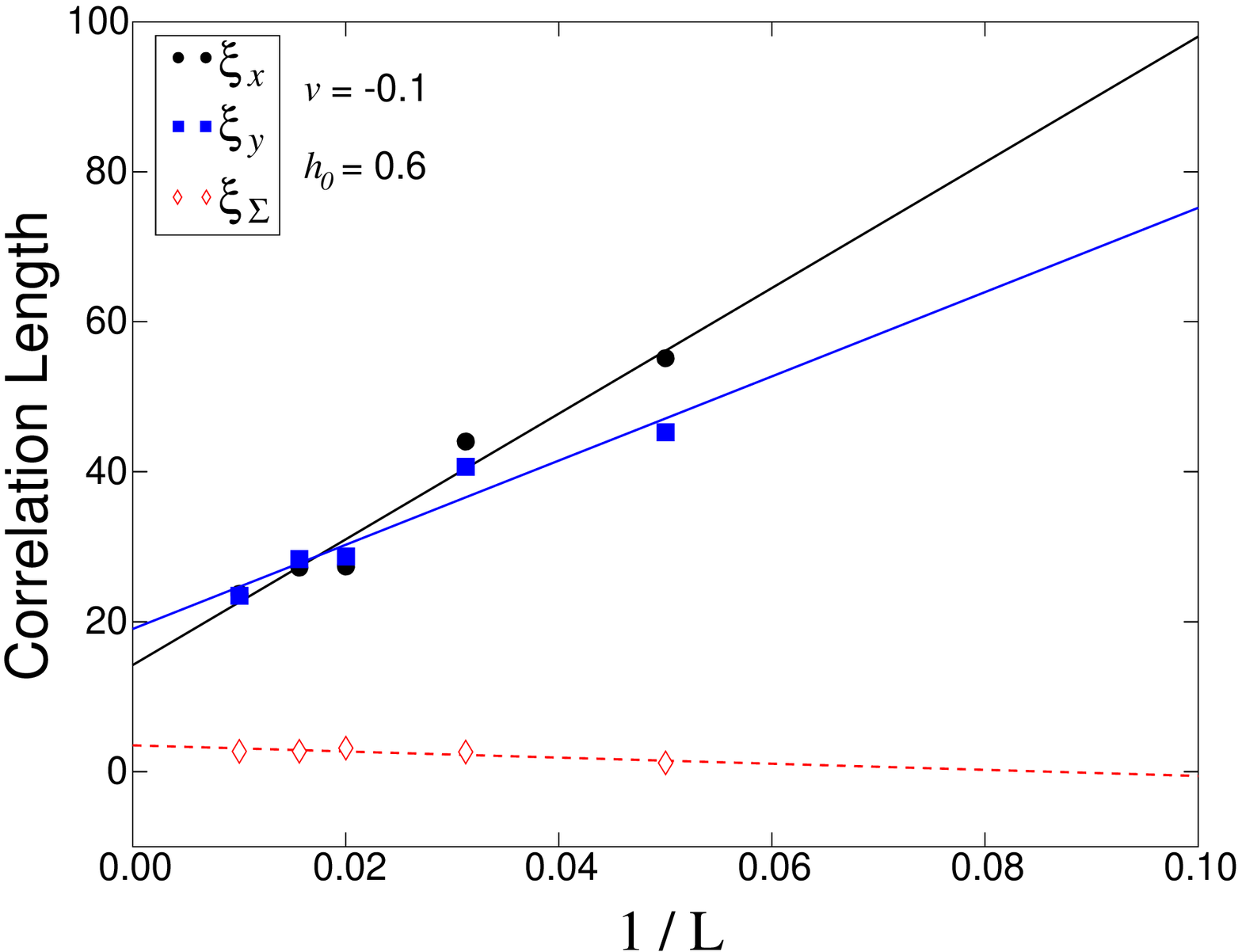}
\caption{Finite size scaling of the correlation lengths $\xi_x$,
$\xi_y$ and $\xi_\Sigma$ using system sizes of $L =
\{20,32,48,64,100\}$ for $v = -0.1$ and $h_0=0.6$. $y$-intercepts
were extracted to approximate $L\rightarrow \infty$.}
\label{fig:fss}
\end{figure}

In order to determine the decay constants associated with $G_\mu(r)$
and $G_\Sigma(r)$ we have performed a discrete Fourier transform of
the disorder averaged correlation functions and fit them to a
Lorentzian in $\vec{k}$-space for each linear system size,
$xy-$interaction and random field variance. The actual correlation
length is assumed to be equal to the width of the Lorentzian. Finite
size scaling was then performed for each value of $v$ and $h_0$, as
shown for $v=-0.1$ and $h_0 = 0.6$ in Fig.~\ref{fig:fss} to extract
approximate infinite system size values of $\xi_\mu$ and
$\xi_\Sigma$.

\begin{figure}[t]
\centering
\includegraphics*[width=3.2in]{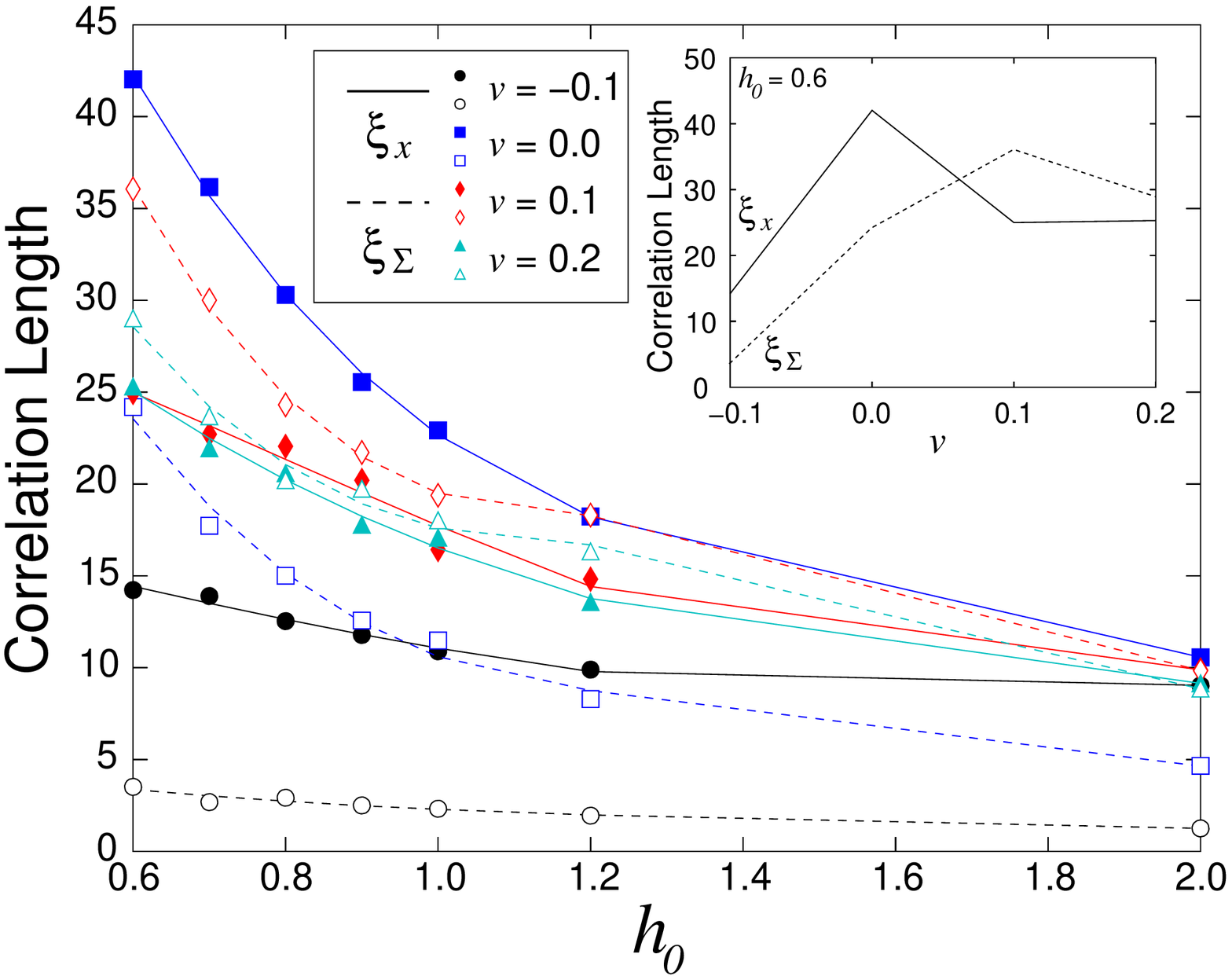}
\caption{The correlation lengths associated with the exponential
decay of $G_x(r)$ and $G_\Sigma(r)$ extracted from finite size
scaled Lorentzian widths in Fourier space.  Color and symbol type
denote different values of $v$, closed symbols with solid lines
refer to $\xi_x$ and open symbols with dashed lines correspond to
$\xi_\Sigma$. The inset shows the two types of correlation lengths
plotted for fixed $h_0$ as a function of $v$. Note that the maximum
values occur at different values of the $xy$ coupling.}
\label{fig:cl}
\end{figure}

The results of the finite size scaling procedure can be seen in
Fig.~\ref{fig:cl} where we plot the decay constants associated with
$G_x$ and $G_\Sigma$ as a function of random field variance $h_0$
for various values of $v$, the error-bars in the fits are on the
order of the symbol sizes.

For a fixed value of $v$, both $\xi_x$ and $\xi_\Sigma$ decrease
monotonically as a function of disorder strength.  The dependence of
correlation lengths on the $xy$ coupling $v$ for a fixed random
field strength is more interesting, as both correlation lengths are
non-monotonic functions of $v$.  Changing $v$ from $-0.1$ to $0.0$,
the correlation length $\xi_x$ increases by almost ten lattice
spacings in the regime of moderate disorder strength.  This increase
can probably be attributed to the fact that for $v = 0.0$, our model
decomposes into two completely uncoupled unidirectional CDWs with
ordering in the $x$ and $y$ direction, respectively.  As disorder
has a weaker influence on a unidirectional CDW as compared to a
checkerboard CDW, this dependence of the correlation length on the
value of $v$ is plausible. A further increase of $v$ to positive
values leads to a reduction of $\xi_x$.

As $v$ controls the degree of stripe order, it is expected that the
correlation length for the Ising-like order parameter increases from
only a few lattice spacings for $v = -0.1$ to 20 to 40 lattice
spacings for $v = 0.1$ in the regime of moderate disorder strength.
It is important to note that for $v=-0.1$ and $v=0.0$, Ising
correlations are exclusively due to disorder fluctuations.  When $v$
approaches zero coming from negative values, fluctuations of the two
fields $\Phi_x$ and $\Phi_y$ are increasingly independent, and Ising
correlations increase. When moving further into the stripe regime,
we find $\xi_\Sigma(v=0.1) > \xi_\Sigma(v=0.2)$ for all values of
$h_0$. This can be attributed to the sharpening of domain walls
between stripey regions with relative orientation $\pi/2$ allowing
them to better accommodate the value of the local random field,
increasing their overall length.

While $\xi_x$ reaches a maximum value at $v = 0$, $\xi_\Sigma$ peaks
at $v = 0.1$ (see inset of Fig.~\ref{fig:cl}). The fact that these
peaks occur at different values of $v$ leads to the interesting
situation that $\xi_\Sigma > \xi_x$ for positive $v$. Hence, with
respect to the analysis of experimental data $\xi_\Sigma > \xi_x$ is
a clear signature for a striped charge order if no random potential
was present.  In the limit of large $v$, the system breaks up into domains 
with either $\Phi_x$ or $\Phi_y$ order and the correlation lengths 
$\xi_x$ and $\xi_\Sigma$ become equal to each other.


\section{Empirical Determination of Stripeness}
\label{sec:eds}

Due to the definition of our effective model for density
fluctuations on the square lattice we have the direct ability to
measure the value of Eq.~(\ref{eq:sigma}) using the ground state
values of the independent CDW order parameters $\Phi_{x,y}$.
However, as we wish to make contact with the STM measurements
discussed in Section~\ref{sec:intro} it would be useful to determine
a method of analyzing $\delta\rho(\vec{r})$ directly which might
expose any underlying local Ising-like correlations that are not
readily apparent either in real or Fourier space.

With this goal in mind, we define an effective local Ising-like
order parameter $\widetilde{\Sigma}(\vec{r})$ at each point in the
sample through the following procedure with more detail provided in an
appendix.  

\textbf{1.} Surround each lattice site $\vec{r}$ with a
$4\times4$ box $\Box_{\vec{r}}$ (depicted in Fig.~\ref{fig:box})
which is ``centered'' at the relative $(1,1)$ point. 

\textbf{2.}
Define a local density-density correlation function which lives in
each $4\times4$ box (the smallest that contains enough information
to resolve the wavevectors $\{\pm \vec{K}_x,\pm\vec{K}_y\}$) that is
arbitrarily assigned to the $(1,1)$ point. The result is $N$
$4\times4$ matrices
\begin{equation}
S_{\Box}(\vec{r},\vec{r}'-\vec{r}'') = \left \langle \delta \rho(\vec{r}')
    \delta\rho(\vec{r}'') \right \rangle_{\vec{r}' - \vec{r}'' \in \Box_{\vec{r}}}
\label{eq:ssquare}
\end{equation}
where $\langle \cdots \rangle_{\vec{r}' - \vec{r}'' \in
\Box_{\vec{r}}}$ indicates an average over all sites $\vec{r}' \in
\Box_{\vec{r}}$ and whose rows and columns are labeled by the $x$
and $y$ components of $\vec{r}'-\vec{r}''$ employing periodic
boundary conditions for the box. The specific form of one of the
matrices contributing to this sum is given in Eq.~(\ref{eq:rhomat}).

\begin{figure}[t]
\centering
\includegraphics*[width=3.2in]{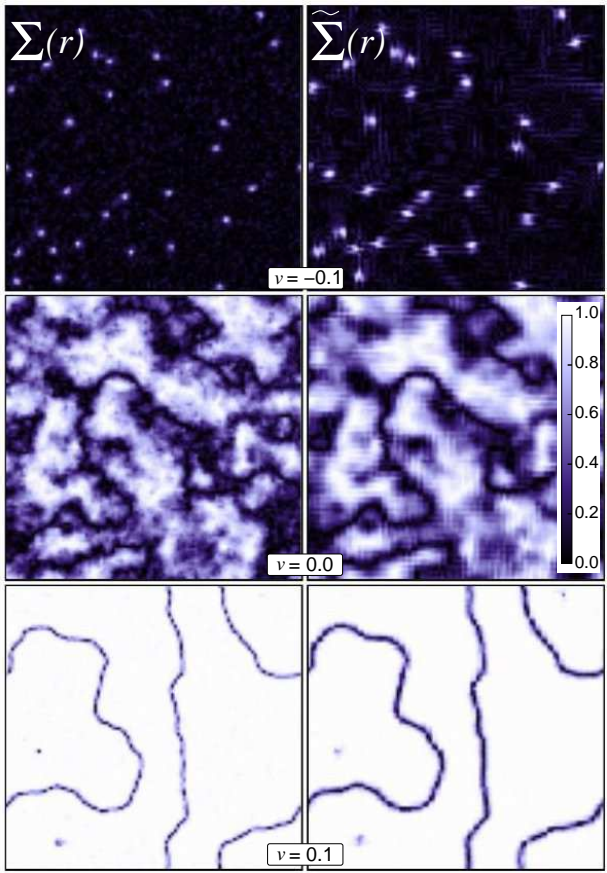}
\caption{A comparison between the direct ($\Sigma$, left-panels) and
effective
         ($\widetilde{\Sigma}$, right-panels) Ising-like order parameters for $h_0 = 1.0$ and
         $v = \{-0.1,0.0,0.1\}$ for a particular realization of disorder in a $100\times100$ sample.}
\label{fig:comp}
\end{figure}

\textbf{3.} Perform a local discrete Fourier transform of
Eq.~(\ref{eq:ssquare}) using only those points in $\Box_{\vec{r}}$
\begin{equation}
S_\Box (\vec{r},\vec{k}) = \frac{1}{16} \sum_{\vec{r'}\in{\Box}_{\vec{r}}}
    S_{\Box}(\vec{r},\vec{r}')\mathrm{e}^{-i\vec{k}\cdot\vec{r}'}
\label{eq:ssquareq}
\end{equation}
for $\vec{k} \in \{\pm\vec{K}_x,\pm\vec{K}_y\}$ at each of the
$L\times L$ boxes. \textbf{4.} Finally, define an effective local
Ising-like order parameter as the difference of local
structure factor amplitudes at $\pm \vec{K}_x$ and $\pm \vec{K}_y$
scaled by their sum in each box
\begin{widetext}
\begin{equation}
\widetilde{\Sigma} (\vec{r}) = \frac{ S_\Box (\vec{r},\vec{K}_x) +
                             S_\Box (\vec{r},-\vec{K}_x) - S_\Box (\vec{r},\vec{K}_y) -
                 S_\Box (\vec{r},-\vec{K}_y)}
                 {S_\Box (\vec{r},\vec{K}_x) + S_\Box (\vec{r},-\vec{K}_x)
                 + S_\Box (\vec{r},\vec{K}_y) + S_\Box (\vec{r},-\vec{K}_y)}.
\label{eq:effsigma}
\end{equation}
\end{widetext}
We may now directly compare Eqs.~(\ref{eq:sigma}) and
(\ref{eq:effsigma}) for different values of $v$ at fixed $h_0$ as
seen in Fig.~\ref{fig:comp}.
The similarity between $\Sigma(\vec{r})$ and
$\widetilde{\Sigma}(\vec{r})$ is striking and shows both environs of
checkerboard order (dark regions, $|\Phi_x(\vec{r})| =
|\Phi_y(\vec{r})|$) and stripe order (light regions,
$|\Phi_x(\vec{r})| \ne |\Phi_y(\vec{r})| = 0$).

The agreement between the left and right panels in
Fig.~\ref{fig:comp} can be further quantized by defining an equal
point correlator
\begin{equation}
\mathcal{Q} = \frac{\Bigl \langle
\Sigma(\vec{r})\widetilde{\Sigma}(\vec{r}) \Bigr \rangle
            - \Bigl \langle \Sigma(\vec{r}) \Bigr \rangle
              \Bigl \langle \widetilde{\Sigma}(\vec{r}) \Bigr \rangle}
              {\sigma_\Sigma \sigma_{\widetilde{\Sigma}}}
\label{eq:Q}
\end{equation}
where $\sigma_{\Sigma}$ is the standard deviation of
$\Sigma(\vec{r})$
\begin{equation}
\sigma_\Sigma = \sqrt{\langle \Sigma^2(\vec{r})\rangle
            - \langle \Sigma(\vec{r})\rangle^2}.
\label{eq:std}
\end{equation}
Using this definition we find values for $\mathcal{Q}$ of $0.62$,
$0.92$ and $0.71$ for $h_0 = 1.0$ and $v = -0.1,0.0$ and $0.1$. The
smaller correlations for $v = \pm 0.1$ are due to the fact that
$\Box_{\vec{r}}$ cannot be constructed symmetrically about the site
$\vec{r}$ as we need to resolve the specific wavevectors $\{\pm
\vec{K}_x, \pm \vec{K}_y\}$ and thus small regions with rapid
changes in $\Phi_{x,y}$ (sharp domain walls) will be poorly
reproduced by the effective field $\widetilde{\Sigma}$.

\begin{figure}[t]
\centering
\includegraphics*[width=3.2in]{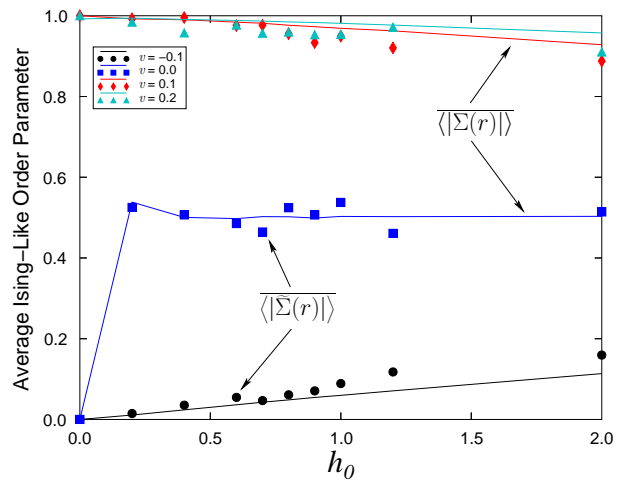}
\caption{A comparison of the disorder averaged magnitudes of
$\Sigma$ and
         $\widetilde{\Sigma}$ for a $100\times100$ sample as a function of the variance
     of the random field.  Finite size scaling appears to have little effect on these
     results.  Symbols show the value of
     $\overline{\langle |\widetilde{\Sigma}(\vec{r})| \rangle}$ while lines refer to
     $\overline{\langle |\Sigma(\vec{r})| \rangle}$.}
\label{fig:avsig}
\end{figure}

A further comparison can be made by examining the disorder averaged
values of the magnitudes of the direct and effective Ising-like
order parameters,
\begin{eqnarray}
\overline{|\Sigma|} &=& \overline{\langle |\Sigma(\vec{r})| \rangle}  \nonumber \\ 
\overline{|\widetilde{\Sigma}|} &=& \overline{\langle
|\widetilde{\Sigma}(\vec{r})| \rangle} \label{eq:aviop}
\end{eqnarray}
which are shown in Fig.~\ref{fig:avsig} for various values of $v$ as
a function of disorder.
Again we observe significant agreement between the direct and
effective Ising-Like order parameters, now having averaged over 100
realizations of disorder.  At $h_0 = 0.0$ we recover the results that
in the disorder free theory $\overline{|\Sigma|}$ should be
identically zero for $v \le 0$ and equal to one for $v > 0$.
Increasing disorder causes a smooth increase in unidirectional order
for $v < 0$ and a reduction for $v > 0$ with the effective order
parameter having a slightly larger dependence on $h_0$. For $v = 0.0$, 
the magnitude of Ising-like order quickly jumps to a value near $0.5$.

The concurrence between $\Sigma(\vec{r})$ and the effective object
$\widetilde{\Sigma}(\vec{r})$ inferred from the scaled difference in
local structure factor peaks amplitudes at wavevectors $\pm
\vec{K}_x$ and $\pm \vec{K}_y$ is not surprising in the context that
they are both calculated from the same underlying complex fields
$\Phi_{x,y}$ in the condensed phase.  However, the utility of
Eq.~(\ref{eq:effsigma}) becomes apparent when considering the
plethora of experimental STM spectra where only the LDOS is measured
and the underlying order is a topic of hot debate. The current
analysis of such spectra involves performing a discrete Fourier
transform over the entire field of view. In real materials, disorder
plays an important role, and short density-density correlation lengths
are often observed.  Therefore, such a \emph{global} Fourier
transform discards a large amount of relevant local information
which could in principle be used to probe any hidden electronic
structure.


\section{The Uncoupled Theory}
\label{sec:uct}

In the limit of large disorder a number of results can be explained
for the uncoupled theory with $v = 0$.

Upon examination of the various correlation lengths in
Fig.~\ref{fig:cl} it is apparent that $\xi_x$ is almost twice as
large as $\xi_\Sigma$ for $v=0$. This can be understood with the
help of an approximate analytical argument: for $v=0$, the lattice model
Eq.~(\ref{eq:FL}) reduces to two uncoupled unidirectional CDWs in
the $x$- and $y$-directions. Concentrating on the numerator of the Ising
order parameter Eq.~(\ref{eq:sigma}) for the moment, in the case
$v=0$ the correlation function of $\Sigma$ is then proportional to
$\left\langle |\Phi_x(r)|^2 |\Phi_x(0)|^2\right\rangle$. If
fluctuations of $\Phi_x$ were described by a simple Gaussian theory,
this would imply $\xi_x = 2 \xi_\Sigma$, which indeed is
approximately found in Fig.~\ref{fig:cl}.  

A similar approach can be used to account for the saturation of
$\overline{|\Sigma|}$ and  $\overline{|\widetilde{\Sigma}|}$ to
$0.5$ for $v=0$ as seen in Fig.~\ref{fig:avsig}. Again if we assume
uncoupled  fluctuations in $\Phi_{x,y}$ described by a normal
distribution with mean $0$ and variance $\sigma_\Phi^2$ we can
directly calculate the expectation value of $\Sigma$ by performing
integrals in polar coordinates
\begin{equation}
\left\langle
\frac{(|\Phi_x|^2-|\Phi_y|^2)^2}{(|\Phi_x|^2+|\Phi_y|^2)^2} \right
\rangle = \frac{1}{2}.
\end{equation}


\section{Conclusions}

This paper has characterized the correlations in a disordered CDW
state on the square lattice as a function of the parameter $v$ in
$\mathcal{F}_\Phi$ (which measures the degree of unidirectionality
of the CDW order, $v<0$ corresponding to checkerboard states), and
the random field strength $h_0$. We introduced a number of
diagnostics to study the nature of the disordered state: the
correlation lengths, $\xi_\mu$, of the CDW order, the correlation
length $\xi_\Sigma$ of the Ising order associated with the
uni-directionality, the on-site amplitudes $G_\mu (0)$, $G_\Sigma
(0)$, of these orders, and also discussed in Section~\ref{sec:eds}
how closely related quantities could be measured directly in
experiments.

Our results are presented in detail in Sections~\ref{sec:results}
and~\ref{sec:eds}, and here we highlight some notable features:\\
({\em i\/}) As is clear from the insets of Fig.~\ref{fig:G}, for
$v<0$, the strength of the Ising order {\em increases\/} with
random-field strength,
while the opposite is true for $v>0$.\\
({\em ii\/}) The correlation length $\xi_\Sigma > \xi_\mu$ only for
$v>0$, and this can serve as a diagnostic for the sign of $v$ in an
analysis of the experiments.\\
({\em iii\/}) We showed that the correlator $S_\Box$ in
Eq.~(\ref{eq:ssquare}) can serve as a very faithful diagnostic of
the structure of $\Sigma$, and this should easily enable us to place
experimental observations in the appropriate parameter space of the
present theory.

A full interpretation of the experiments should away a direct
analysis of the experimental data along the lines proposed here.
Nevertheless, a comparison of the qualitative structure of the
figures presented here (especially Fig~\ref{fig:drhovh}) with {\em
e.g.\/} the STM observations of Ca$_{2-x}$Na$_{x}$CuO$_{2}$Cl$_{2}$
by Hanaguri {\em et al.} does indicate that this system has a value
of $v$ close to zero and likely positive.

For the future, our approach offers an avenue to quantitatively
correlate the STM experiments with neutron scattering. In
particular, after determining the appropriate parameter regime of
$\mathcal{F}_\Phi$ from an analysis of the STM data, the resulting
$\delta \rho (\vec{r})$ configurations can be used as an input to
determining the dynamic spin structure factor, as discussed in
recent works \cite{ssvojta,ribhu}.

While this paper was being completed, we learned of related results
obtained by Robertson {\em et al.} \cite{robertson}.


\acknowledgments The authors thank L.~Bartosch, J.~Hoffman,
A.~Kapitulnik, S.~Kivelson, and R.~Melko for discussions relating
to multiple aspects of this work. In addition, this work benefitted
from valuable discussions with T.~Nattermann. We thank J.~Robertson
for giving us a preview of their related results
\cite{robertson}.This research was supported by NSF Grant
DMR-0537077. A.~D. would like to thank NSERC of Canada for financial
support through Grant PGS D2-316308-2005.  B.~R. acknowledges support 
through the Heisenberg program of DFG.  The computer simulations
were carried out using resources provided by the Harvard Center for
Nanoscale Systems, part of the National Nanotechnology
Infrastructure Network.

\appendix

\section{Local Ising-Like Order Parameter}

This appendix provides more detail on the method used in the
calculation of the effective local Ising-like order parameter
$\widetilde{\Sigma}(\vec{r})$ described in Sec.~\ref{sec:eds}.

\begin{figure}[t]
\centering
\includegraphics[width=2.3in]{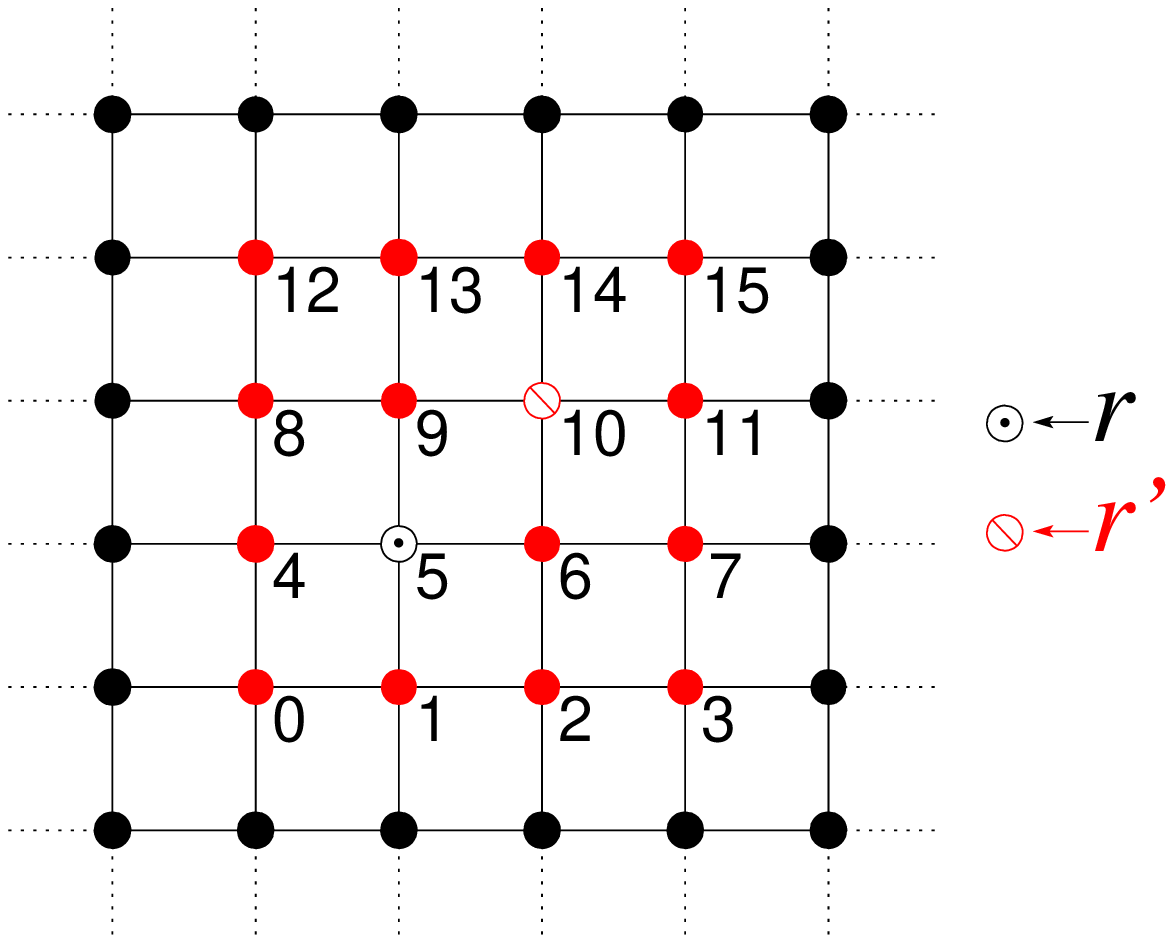}
\caption{One of $N$ $4\times4$ boxes used in the calculation of 
$S_{\Box}(\vec{r},\vec{r}'-\vec{r}'')$.  Here, $\Box_{\vec{r}}$ has relative 
site labels 0 to 15 and is centered at site 5 with $\vec{r}'$ currently located 
at site 10.} 
\label{fig:box}
\end{figure}

Consider Fig.~\ref{fig:box} which shows one of $N$ $4\times4$ boxes
$\Box_{\vec{r}}$ where the sites contained within the box are given
the relative labels 0 through 15 and it is ``centered'' at the
$(1,1)$ point here labelled 5.

For the particular case shown in Fig.~\ref{fig:box} with $\vec{r}' =
\vec{r}'_{10}$ the matrix which contributes to the average
$S_{\Box}(\vec{r},\vec{r}'-\vec{r}'')$ in Eq.~(\ref{eq:ssquare}) is
written out explicitly as
\begin{widetext}
\begin{equation}
S_{\Box}(\vec{r},\vec{r}'-\vec{r}'') = \cdots +
\frac{\dr{10}}{16}\left( \begin{array}{cccc}
\dr{14}+\dr{6} & \dr{5}+\dr{7}+\dr{13}+\dr{15} & 2(\dr{4}+\dr{12}) & \dr{5}+\dr{7}+\dr{13}+\dr{15} \\
2\dr{2} & 2(\dr{1}+\dr{3}) & 4\dr{0} & 2(\dr{1}+\dr{3}) \\
\dr{14}+\dr{6} & \dr{5}+\dr{7}+\dr{13}+\dr{15} & 2(\dr{4}+\dr{12}) & \dr{5}+\dr{7}+\dr{13}+\dr{15} \\
\dr{10} & \dr{9}+\dr{11} & 2\dr{8} & \dr{9}+\dr{11}
\end{array} \right) + \cdots
\label{eq:rhomat}
\end{equation}
\end{widetext}
where we have employed the shorthand notation $\dr{i} \equiv
\delta\rho(\vec{r}'_{i})$. After performing the average over all
$\vec{r}'-\vec{r}'' \in \Box_{\vec{r}}$ i.e.~calculating all 16
matrices at each site, $S_{\Box}(\vec{r},\vec{r}'-\vec{r}'')$ is
Fourier transformed over the relative coordinates in the box using
Eq.~(\ref{eq:ssquareq}) to give $S_{\Box}(\vec{r},\vec{k})$. This
expression may then be evaluated at $\pm \vec{K}_x$ and $\pm
\vec{K}_y$ using Eq.~(\ref{eq:effsigma}) to give the effective Ising-like
order parameter $\widetilde{\Sigma}(\vec{r})$.

\end{document}